\documentclass[%
 reprint,
 amsmath,amssymb,
 aps,
]{revtex4-2}

\setcitestyle{square, numbers,compress}
\usepackage{graphicx}
\usepackage{wrapfig}
\usepackage{dcolumn}
\usepackage{bm}
\usepackage{amsmath}
\usepackage{amssymb}
\usepackage{gensymb}

\begin{document}

\title{Constructing a Quantum Twisting Microscope: Design Insights and Experimental Considerations}

\author{Sayanwita Biswas}
\thanks{These two authors contributed equally} 

\author{Ranjani Ramachandran}
\thanks{These two authors contributed equally} 

\author{Patrick Irvin}

\author{Jeremy Levy}
\affiliation{Department of Physics and Astronomy, University of Pittsburgh, Pittsburgh, Pennsylvania 15260, USA}
\affiliation{Pittsburgh Quantum Institute, Pittsburgh, Pennsylvania 15260, USA}

\date{\today}

\begin{abstract}
We report the details of construction and testing of a Quantum Twisting Microscope (QTM), a recently developed scanning probe instrument \cite{Inbar2023-pp} that enables twist-angle-dependent electronic measurements on layered materials. Our implementation is based on a commercial atomic force microscope (Nanosurf Easyscan 2) whose open geometry beneath the scan head allows integration of the rotation and translation stages required for QTM operation. We describe the complete fabrication process including tip preparation by focused ion beam deposition and graphite transfer, custom stage assembly with integrated rotation capability, and multi-step alignment procedures. To validate the instrument, we perform conductance measurements between graphite layers as a function of twist angle, observing clear 60-degree periodicity consistent with the hexagonal lattice symmetry and conductance enhancements near the commensurate twist angles of $21.8\degree$ and $38.2\degree$. These results confirm the instrument's ability to resolve crystallographic twist-angle-dependent transport features. By providing detailed construction and operational guidelines, we aim to make QTM technology accessible to research groups with standard AFM infrastructure, enabling investigations of twist-angle-dependent phenomena in van der Waals materials, complex oxide heterostructures, and chiral systems.
\end{abstract}

\maketitle


\section{Introduction}

\begin{figure*}
    \centering
    \includegraphics[width=\textwidth]{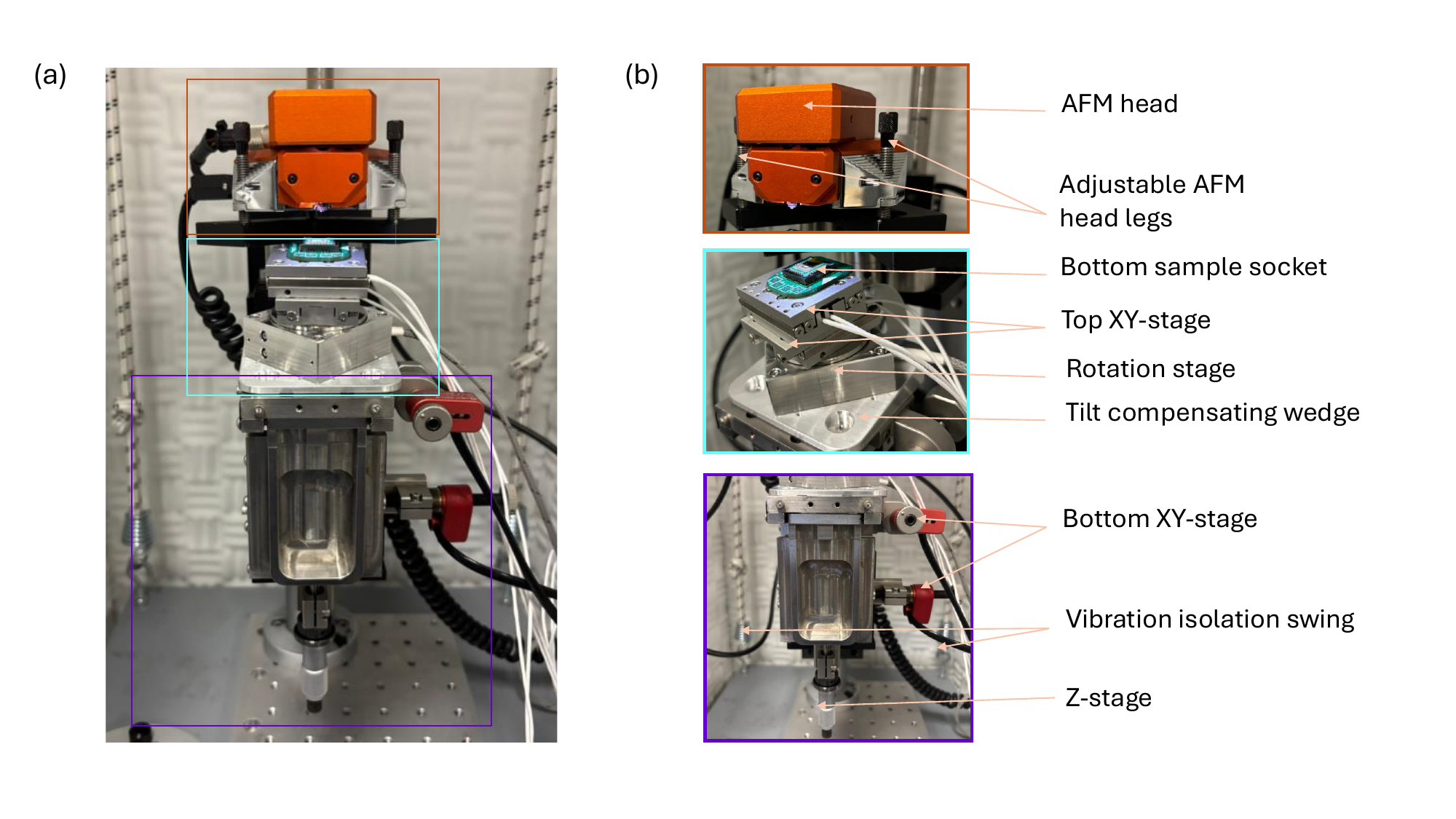}
    \caption{Camera images of QTM setup: (a) Full instrument image showing the complete assembly. (b) Detailed view of different instrument components including the AFM head, translation stages, rotation stage, and sample mount.}
    \label{fig:afm}
\end{figure*}

The advent of two-dimensional materials has opened new frontiers in condensed matter physics, particularly through the discovery of emergent phenomena in twisted heterostructures. When adjacent layers of materials such as graphene are rotationally misaligned, they form moir\'e superstructures that dramatically alter their electronic behavior \cite{Ribeiro-Palau2018-sa, Cao2018-ze, Cao2018-mo}. Twisting bilayer graphene \cite{Chari2016-gl}, in particular, has revealed remarkable properties including unconventional superconductivity and correlated insulating states at specific ``magic angles.'' The ability to engineer electronic properties through rotational alignment has motivated the broader field of ``twistronics'' \cite{Yang2020-es}, extending to transition metal dichalcogenides, complex oxides, and other layered systems.

The recently developed Quantum Twisting Microscope (QTM) \cite{Inbar2023-pp} addresses a central experimental challenge in this field: the need for precise, continuous control over the twist angle between layered materials while simultaneously probing their electronic properties. The QTM uses a short pyramid-shaped tip covered with a two-dimensional material to form tunneling junctions with a flat substrate. Unlike conventional scanning tunneling microscopes, the QTM can access momentum-resolved tunneling information by varying the twist angle, enabling measurements that reveal the evolution of electronic states as a function of interlayer rotation. This capability makes it a powerful tool for investigating a wide range of twist-angle-dependent phenomena, from moir\'e-engineered band structures to chiral transport effects.

In this article, we present the construction of a QTM based on a commercial AFM platform whose open geometry makes it well suited for the required modifications. We detail the technical challenges encountered during development and demonstrate the instrument's capabilities through twist-angle-dependent conductance measurements between graphite layers. Our measurements reveal the characteristic 60-degree symmetry expected from the hexagonal crystal structure and show conductance enhancements near the commensurate twist angles of $21.8\degree$ and $38.2\degree$ \cite{Koren2016-ba, Chari2016-gl, Bistritzer2010-md}, establishing the microscope's ability to resolve crystallographic transport features. By providing detailed construction and operational guidelines, we aim to lower the barrier for other research groups to adopt QTM technology.

\section{Instrument Design, Construction, and Operation}

The QTM consists of three primary functional components: a modified atomic force microscope (AFM) serving as the scanning and positioning system, custom-fabricated QTM tips with specific geometric requirements, and a specialized stage assembly incorporating translation and rotation capabilities. Figure \ref{fig:afm} shows the complete assembled instrument and its key components. We describe each component in detail, emphasizing critical design considerations and solutions to technical challenges, followed by the operational procedures required for successful measurement.

\subsection{AFM Platform Selection and Modification}

We based our QTM on a Nanosurf Easyscan 2 AFM. Unlike many commercial AFMs (e.g., the Cypher AFM), the Easyscan 2 head can be elevated and mounted on any post with adjustable height, leaving substantial open space beneath it. This open geometry is essential for integrating the custom rotation and translation stages required for QTM operation. The adjustable legs also allow precise control of the angle between the cantilever and the substrate, which proves critical for accommodating the shorter QTM tip profile as discussed below. The Easyscan 2 has been discontinued by Nanosurf, but the two key design requirements for QTM construction (a tip-scanning head with open access beneath it, and adjustable legs that allow the tilt angle between cantilever and sample to be changed) are met by instruments from several manufacturers across a wide range of price points.

The AFM operates in contact mode with continuous twist motion, requiring modifications to standard AFM operation procedures. The cantilever holder was modified and custom tips were fabricated to ensure electrical connectivity while maintaining mechanical stability during rotation. Additionally, the feedback control parameters were optimized for maintaining constant force during twist operations rather than standard topographic scanning.

\begin{figure*}
    \centering
    \includegraphics[]{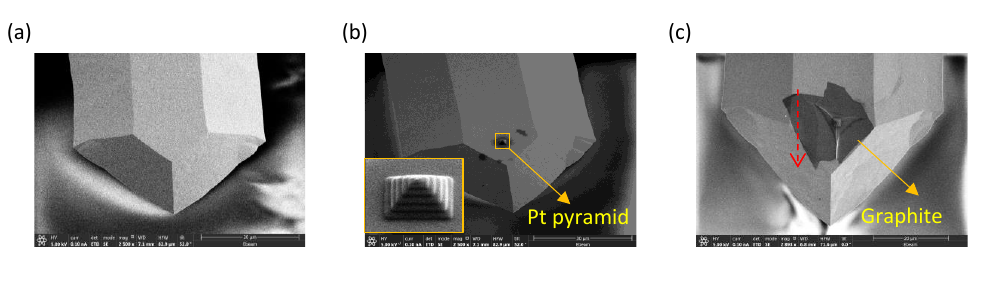}
    \caption{SEM images showing the QTM tip fabrication process: (a) Tipless cantilever after gold deposition showing the metallic coating. (b) Cantilever after FIB deposition of platinum pyramid with inset showing a magnified view of the pyramid structure. (c) Completed QTM tip after transfer of graphite layer onto the pyramid, creating the functional tip surface.}
    \label{fig:tip making}
\end{figure*}

\subsection{QTM Tip Fabrication}

The fabrication of QTM tips requires precise control over geometry, materials, and electrical properties. 
We provide detailed step-by-step insights and discuss critical practical challenges to facilitate anyone interested in the setting up a QTM.

We begin with tipless cantilevers from Nanosensors, specifically selected for their high spring constant of 48 N/m (length 220~$\mu$m and width 40~$\mu$m). The higher spring constant compared to standard AFM cantilevers (typically $0.1-10$~N/m) is essential for maintaining stable contact during continuous twist operations while preventing damage to the delicate 2D material layers.

Electrical contact is established through e-beam evaporation of a 4~nm chromium adhesion layer followed by 100~nm of gold onto the cantilever (Fig.~\ref{fig:tip making}(a)). The chromium layer is critical for ensuring good adhesion of the gold to the silicon nitride cantilever surface. The deposited gold film provides low electrical resistance while maintaining mechanical flexibility of the cantilever.

The platinum pyramid is fabricated using focused ion beam (FIB) deposition with carefully optimized parameters: acceleration voltage of 30~keV, ion beam current of 10~pA, and dwell time of 800~ns. Critical to successful deposition is proper grounding of the cantilever and continuous drift correction throughout the process. The resulting pyramid has a $2 \times 2$~$\mu$m base and stands $1.5-2.0~\mu$m tall (Fig. \ref{fig:tip making}(b)).

\begin{figure}
    \centering
    \includegraphics[width=0.8\linewidth]{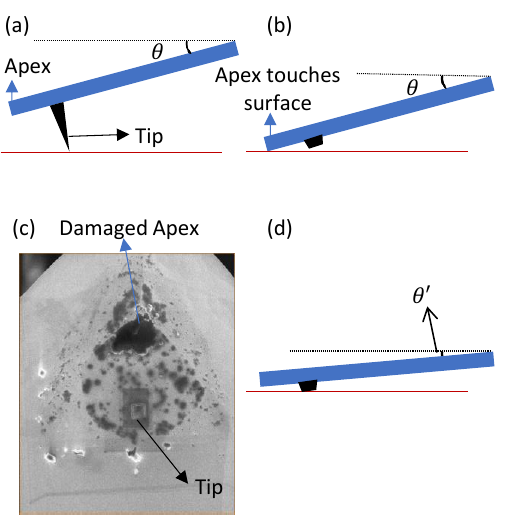}
    \caption{Critical tip height considerations for QTM operation: (a) Schematic of conventional AFM tip touching the surface at the standard angle of $\theta \approx 9-12\degree$. (b) QTM tip with insufficient height causing cantilever apex to contact surface before tip engagement. (c) SEM image showing cantilever damage from apex contact when tilt angle is not properly optimized. (d) Optimized configuration with adjusted tilt angle ($\theta' < \theta$) ensuring only the tip contacts the surface.}
    \label{fig:apextouch}
\end{figure}

The pyramid height is a critical parameter that must be carefully controlled. The original QTM publication \cite{Inbar2023-pp} reports pyramid heights of approximately $1.2-1.6~\mu$m; in our implementation we find an optimal range of $1.5-2.0~\mu$m, 
with the difference likely arising from the tilt angle specific to each setup, which will be discussed in the next section. If the pyramid is too tall ($> 2.5~\mu$m), the subsequently transferred membrane will not form the required smooth tent-like structure, compromising electrical contact and mechanical stability. Conversely, if the pyramid is too short ($< 1.5~\mu$m), the cantilever apex may contact the sample surface before the tip engages [Figure \ref{fig:apextouch}], preventing proper operation.

The final tip fabrication step involves transferring graphite onto the platinum pyramid using a modified PDMS-based transfer technique \cite{Jayasena2015-bn}. Several factors determine transfer success. The graphite flakes must have appropriate dimensions to maximize contact with the flat cantilever surface surrounding the pyramid without extending onto the sloping sides. Flakes with insufficient contact area or those extending onto slopes exhibit poor adhesion. The optimal flake thickness ranges from 10 to 50~nm; thinner layers tend to wrinkle while thicker layers may not conform properly to the pyramid shape.

The transfer procedure requires careful control. The cantilever is mounted on a silicon wafer with the tip facing upward. Using a standard van der Waals transfer stage, the PDMS stamp carrying the graphite flake is brought into contact with the pyramid. Success rates improve significantly when the glass slide holding the PDMS is slightly tilted, allowing gradual contact starting from the back end of the cantilever and progressing toward the apex (indicated by the red arrow in Fig. \ref{fig:tip making}(c)). Heating the cantilever to $50\degree$C during transfer enhances adhesion through improved van der Waals bonding.

\subsection{Custom Stage Assembly}

The QTM stage assembly addresses several unique requirements not present in conventional AFMs, including precise tilt compensation, multiple translation stages for alignment, and rotation capability while maintaining the center of rotation.

\begin{figure}
    \centering
    \includegraphics[width=1\linewidth]{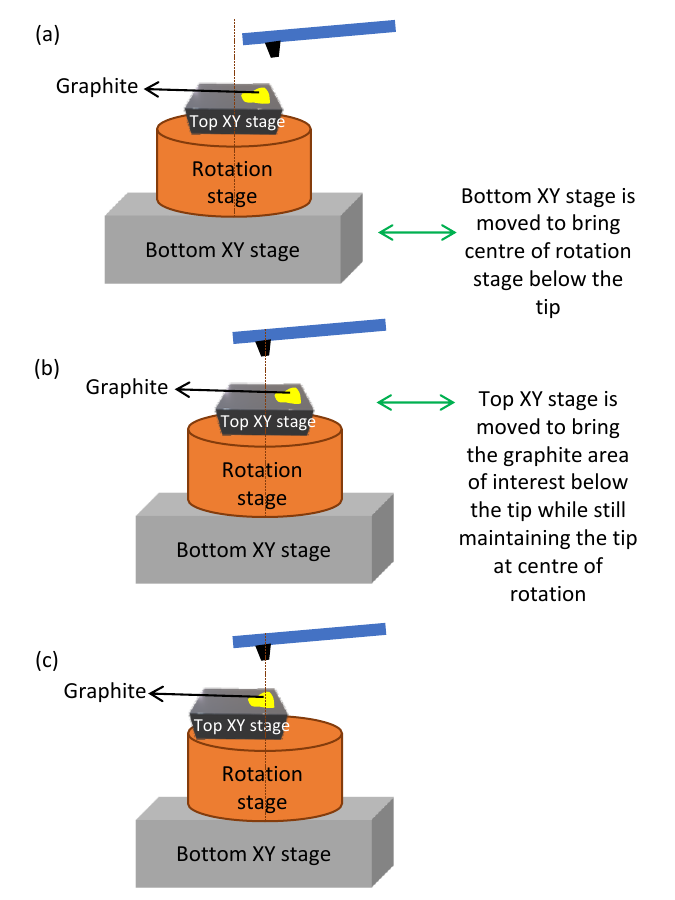}
    \caption{Stage alignment procedure for centering the area of interest at the rotation axis: (a) Initial configuration showing misalignment between tip, graphite area of interest, and rotation stage center (indicated by red dotted line). (b) After bottom XY stage adjustment, the tip is aligned with the rotation center. (c) Final configuration after top XY stage adjustment brings the graphite area to the rotation center, enabling twist-angle measurements without lateral displacement.}
    \label{fig:stage movement}
\end{figure}

A fundamental challenge in QTM design stems from the geometry of commercial AFM heads, which position cantilevers at angles of $9-12\degree$ relative to the sample surface. While appropriate for conventional sharp AFM tips, this angle causes the cantilever apex to contact the sample before the shorter QTM tip can engage (Fig. \ref{fig:apextouch}). We developed a two-stage solution to this problem. First, we incorporated an 8-degree wedge into the stage assembly that reduces the effective tilt angle between the cantilever and sample. Second, we use the three adjustable legs of the AFM head for fine-tuning to achieve optimal alignment. This adjustment must ensure that neither the cantilever apex nor the cantilever mounting clip contacts the sample during operation.

\begin{figure*}
    \centering
    \includegraphics[]{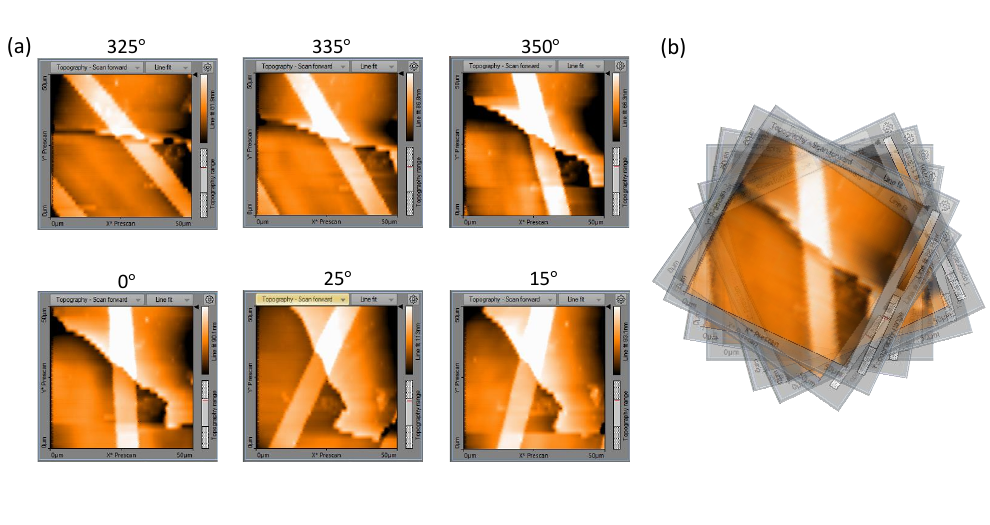}
    \caption{Verification of rotation stage alignment: (a) AFM topography scans acquired at different rotation angles showing consistent feature positions. (b) Superposition of all rotation angles demonstrating that the rotation center is precisely located at the middle of the scan area, confirming successful alignment.}
    \label{fig:centering}
\end{figure*}

The stage assembly incorporates two sets of XY translation stages with distinct functions. The bottom translation stages position the rotation stage axis directly beneath the tip (Fig. \ref{fig:stage movement}(a) and \ref{fig:stage movement}(b)). This alignment is critical for maintaining the tip position over the area of interest during rotation. Coarse alignment uses the AFM's optical system, while fine alignment is achieved through iterative scanning and position adjustment until the rotation center coincides with the scan area center (Fig. \ref{fig:centering}). The top translation stages (Xeryon XLS), mounted above the rotation stage (Xeryon XRT-U), precisely locate the flat sample  relative to the curved sample attached to the QTM tip, without disturbing the tip-rotation axis alignment (Fig. \ref{fig:stage movement}(c)).



The flat sample consists of exfoliated graphite transferred onto a silicon chip with a mesa following established procedures and gold bondpads are added for electrical contact\cite{Inbar2023-pp}. The silicon wafer is mounted on a chip carrier and the flat sample is grounded using aluminum wirebonds to the gold bondpads. The entire assembly design ensures no component extends below the cantilever clip plane, preventing mechanical interference from the cantilever clip and other parts of the AFM head at all rotation angles during measurement. This mechanical interference is a common hurdle that leads to unstable contact between the curved and flat samples. To alleviate this issue, our earlier iterations used a tapered post between the top translational stage and the tip, reducing the area directly beneath the tip to the minimum required for the chip socket.

While it is possible to obtain AFM scans of the topography and even measure some current flow between tip and flat sample at this stage, extensive vibration isolation is critical during continuous rotation measurements, where even small vibrations can cause tip-sample disengagement or introduce artifacts in the conductance measurements. The complete QTM assembly is mounted on a vibration isolation swing inside an acoustic enclosure.

\subsection{Alignment and Measurement Procedures}

Here we describe the QTM alignment procedure. The alignment process begins with mounting the prepared QTM tip in the AFM head and positioning the sample on the upper translation stage. Using the AFM's optical camera, the rotation stage is roughly centered beneath the tip using the bottom translation stages. The wedge angle and AFM leg heights are adjusted to achieve approximately parallel alignment between the cantilever and sample surface.

Fine alignment proceeds through an iterative process. The tip is brought into contact with the sample using standard AFM approach procedures with reduced setpoint force ($10-20$~nN) to prevent tip damage. A series of topography scans are acquired at different rotation angles (typically 0$\degree$, 45$\degree$, 90$\degree$, and 135$\degree$). The scanned images are analyzed to determine the offset between the rotation center and scan center. The bottom translation stages are adjusted to minimize this offset. This process is repeated until the rotation center falls within 100~nm of the scan center (Fig. \ref{fig:centering}).

Once alignment is complete, twist-angle-dependent measurements proceed by first positioning the tip over the desired measurement location using the top translation stages. Contact is established with setpoint force sufficient to ensure stable electrical contact without damaging the graphite layers ($20-50$~nN). 

At this point, it is usually straightforward to observe current flow between the tip and the flat sample. However, a common issue is that the measured conductance, $G$, can remain relatively low (below 1 $\mu$S for our experiments with two layers of graphite), leading to noisy data that is not reproducible as a function of twist angle $\theta$. This highlights the importance of good vibration isolation and clean contact surfaces between the tip layer and flat layer. For the latter, we recommend keeping the tip layer as pristine as possible by avoiding any contact-mode scanning and polymer residue from transfer. Once this is ensured, the precise tip-approach setpoint or slight off-center rotation is typically not critical. Instead, careful tuning of the tilt angle using the AFM legs is required to maintain stable contact. We recommend keeping the tip-sample contact while rotating multiple times at a clean location until the van der Waals coupling stabilizes and $G$ jumps to few microsiemens, as expected for graphite.


\section{Results and Validation}

To establish QTM functionality, we performed twist-angle-dependent conductance measurements between graphite-coated tips and graphite substrates. These measurements serve as a benchmark due to the well-understood electronic structure of graphite and the clear signatures expected from twist-angle modulation.

The bottom graphite sample was electrically grounded while current was measured through the tip graphite layer. An AC excitation of 15~mV at 13 Hz on top of a 40 mV DC bias voltage was applied to the QTM tip, and the rotation stage was continuously rotated at $2\degree~\mathrm{min}^{-1}$ while simultaneously recording the current between the tip to bottom sample. The conductance was calculated from the measured current and applied voltage.

\begin{figure}
    \centering
    \includegraphics[width=1\linewidth]{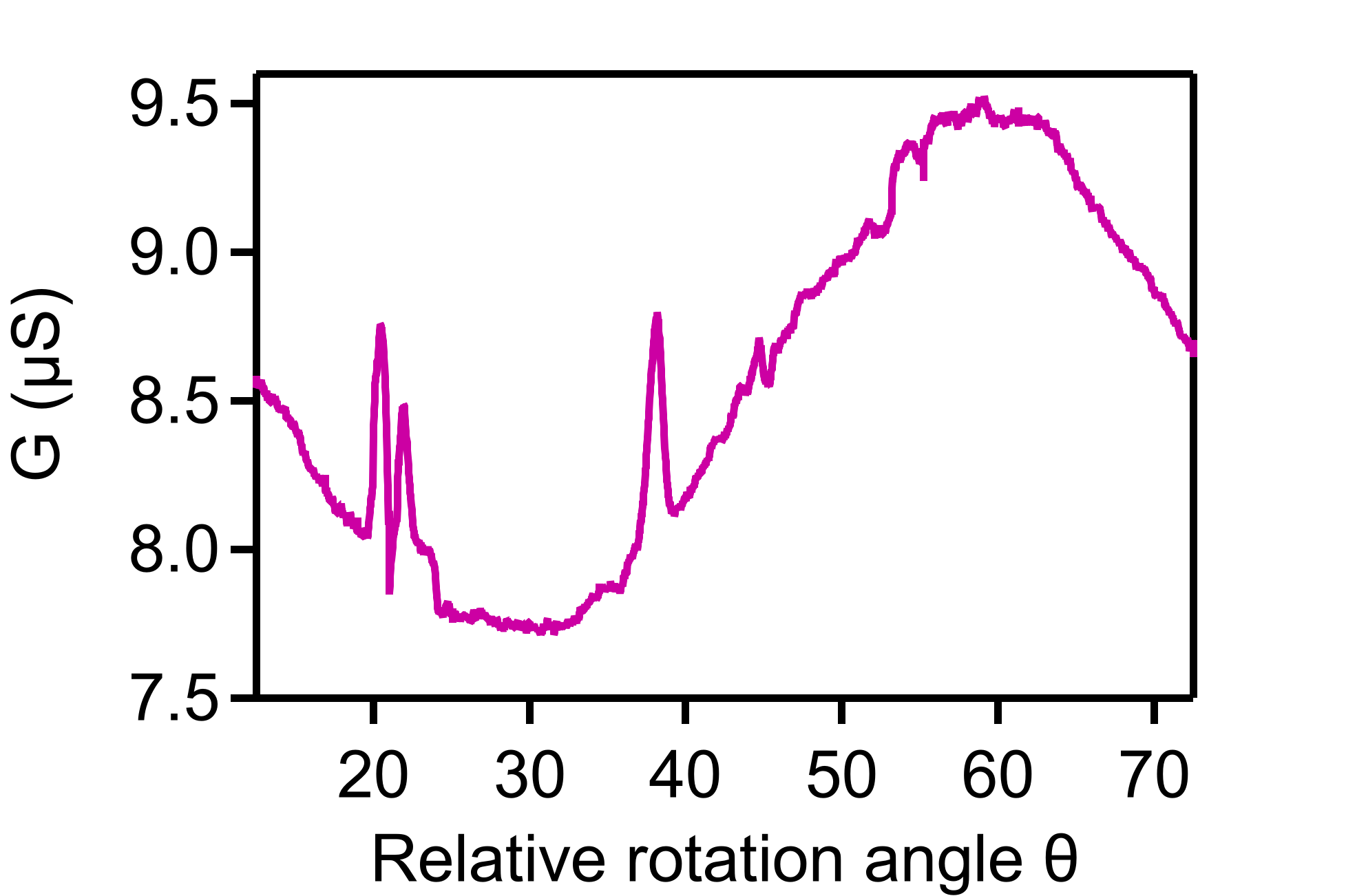}
    \caption{Conductance as a function of relative twist angle $\theta$ between two graphite layers, showing 60-degree periodicity consistent with hexagonal lattice symmetry. Enhanced conductance is observed near the commensurate angles of $21.8\degree$ and $38.2\degree$, corresponding to special stacking configurations that enable resonant interlayer tunneling.}
    \label{fig:result}
\end{figure}

Figure \ref{fig:result} shows the measured conductance, $G$, as a function of relative twist angle between the graphite layers. The data exhibits clear 60-degree-periodicity, consistent with the hexagonal symmetry of the graphite crystal lattice. This periodicity provides strong validation that the QTM is measuring the effect of crystallographic twist rather than artifacts from mechanical motion or electrical contact variations.

Beyond the overall periodicity, we observe distinct conductance enhancements at specific angles. The observed peak positions near $22\degree$ and $38.19\degree$ are consistent with two of the theoretically predicted large-angle commensurate configurations for twisted graphite layers. Bistritzer and MacDonald \cite{Bistritzer2010-md} identified a set of commensurate twist angles ($13.2\degree$, $21.8\degree$, $27.8\degree$, $32.2\degree$, $38.2\degree$, and $46.8\degree$) at which partial overlap of the Fermi surfaces of adjacent layers enables resonant interlayer tunneling. Our observed peaks are consistent with the predicted commensurate angles at $21.8\degree$ and $38.2\degree$.

The enhanced conductance at commensurate angles arises from momentum conservation in interlayer tunneling. At generic twist angles, the Fermi circles of adjacent layers do not overlap in momentum space, suppressing direct tunneling. At commensurate angles, however, partial Fermi surface overlap enables resonant tunneling pathways. The distinction between the two observed peaks relates to different tunneling mechanisms \cite{Bistritzer2010-md}: the peak near $21.8\degree$ corresponds to intervalley tunneling between the K point in one layer and the K$'$ point in the adjacent layer, while the peak near $38.2\degree$ corresponds to intravalley tunneling within the same valley. Intravalley processes generally exhibit higher conductance than intervalley processes due to reduced scattering, which may account for the observed difference in peak heights. Although the overlap sites are intervalley for $13.2\degree$,$21.8\degree$ and $32.2\degree$, we only observe the conductance peak at $21.8\degree$. We do not observe the same characteristic conductance peak at $13.2\degree$ and $32.2\degree$ commensurate angles. This result is consistent with the $13.2\degree$, $21.8\degree$ and $32.2\degree$ conductance peaks calculated theoretically to be different by several orders of magnitude \cite{Bistritzer2010-md}. For example, we examine the $13.2\degree$ superlattice, which is about 2.7-times larger than the superlattice at $21.8\degree$ and therefore corresponds to fewer total commensurability sites over the two graphite layer contact area. Consequently, it may be that the relative impact of the overlap sites at $13.2\degree$ is obscured due to room temperature smearing \cite{Luican2011-hd,Kim2013-al}.

The observed 60-degree-periodicity and conductance peaks at the expected commensurate angles demonstrates that our QTM can resolve twist-angle-dependent transport features associated with crystallographic alignment, validating its capability for investigating a range of 2D material systems.


\section{Conclusions}

In this work, we have described the design and implementation of a QTM based on a commercial atomic force microscope platform. The essential hardware requirements are a tip-scanning AFM head with open access beneath it and adjustable legs that allow the cantilever-sample tilt angle to be changed. These constraints are met by wide range of commercially-available instruments. We detail critical experimental insights to help optimize the parameters for all sub-processes like tip fabrication, flake transfer, tip alignment, and stage modification. We validate its functioning by reproducing the expected twist angle dependence between two layers of graphite. 

Beyond the graphite-on-graphite measurements demonstrated here, the QTM is well-suited for investigating a range of layered materials. Recent advances in freestanding complex oxide membranes \cite{Li2022-sa, Francesco-M-Chiabrera-Shinhee-Yun-Ying-Li-Rasmus-T-Dahm-Haiwu-Zhang-Charline-K-R-Kirchert-Dennis-V-Christensen-Felix-Trier-Thomas-S-Jespersen-and-Nini-Pryds2022-jq} have created opportunities to form twisted oxide heterostructures \cite{Pryds2024-wi} in which strong electronic correlations \cite{Imada1998-mp, Lee2015-pb, Spaldin2010-cb} could be tuned via twist angle. Moir\'e superlattices formed by twisted crystals are also inherently chiral \cite{Menichetti2023-ob}, making the QTM a natural platform for investigating chiral-induced spin selectivity (CISS) \cite{Naaman2012-ow, Bloom2024-bb} in solid-state systems, with implications for spintronics \cite{Yang2021-ld} and for understanding spin-selective transport more broadly.

The QTM opens avenues for investigating quantum materials ranging from twisted van der Waals heterostructures to complex oxide interfaces and chiral systems. By demonstrating that a functional QTM can be built from a commercial AFM platform with standard nanofabrication tools, we hope to facilitate broader adoption of this technique across the condensed matter and materials science communities.

\medskip
\textbf{Acknowledgments} \par
We thank Dr.\ Yuze Meng for training on graphite transfer methods. We thank Dr.\ David Snoke, Qiaochu Wan, and Daniel Vaz for providing access to their transfer stage. We thank Dr.\ Shahal Ilani, John Birkbeck, Alon Inbar, and Jiewen Xiao for technical guidance and for providing graphite sample substrates.



\bibliography{QTMpaperpile}

@ARTICLE{Spaldin2010-cb,
  title     = "Multiferroics: Past, present, and future",
  author    = "Spaldin, Nicola A and Cheong, Sang-Wook and Ramesh, Ramamoorthy",
  journal   = "Phys. Today",
  publisher = "AIP Publishing",
  volume    =  63,
  number    =  10,
  pages     = "38--43",
  month     =  "1~" # oct,
  year      =  2010,
  doi       = "10.1063/1.3502547"
}

@ARTICLE{Ribeiro-Palau2018-sa,
  title     = "Twistable electronics with dynamically rotatable heterostructures",
  author    = "Ribeiro-Palau, Rebeca and Zhang, Changjian and Watanabe, Kenji
               and Taniguchi, Takashi and Hone, James and Dean, Cory R",
  journal   = "Science",
  publisher = "American Association for the Advancement of Science (AAAS)",
  volume    =  361,
  number    =  6403,
  pages     = "690--693",
  month     =  "17~" # aug,
  year      =  2018,
  doi       = "10.1126/science.aat6981"
}

@ARTICLE{Cao2018-ze,
  title     = "Correlated insulator behaviour at half-filling in magic-angle
               graphene superlattices",
  author    = "Cao, Yuan and Fatemi, Valla and Demir, Ahmet and Fang, Shiang and
               Tomarken, Spencer L and Luo, Jason Y and Sanchez-Yamagishi,
               Javier D and Watanabe, Kenji and Taniguchi, Takashi and Kaxiras,
               Efthimios and Ashoori, Ray C and Jarillo-Herrero, Pablo",
  journal   = "Nature",
  publisher = "Springer Science and Business Media LLC",
  volume    =  556,
  number    =  7699,
  pages     = "80--84",
  month     =  "5~" # apr,
  year      =  2018,
  doi       = "10.1038/nature26154"
}

@ARTICLE{Cao2018-mo,
  title     = "Unconventional superconductivity in magic-angle graphene
               superlattices",
  author    = "Cao, Yuan and Fatemi, Valla and Fang, Shiang and Watanabe, Kenji
               and Taniguchi, Takashi and Kaxiras, Efthimios and
               Jarillo-Herrero, Pablo",
  journal   = "Nature",
  publisher = "Springer Science and Business Media LLC",
  volume    =  556,
  number    =  7699,
  pages     = "43--50",
  month     =  "5~" # apr,
  year      =  2018,
  doi       = "10.1038/nature26160"
}

@ARTICLE{Bloom2024-bb,
  title     = "Chiral induced spin selectivity",
  author    = "Bloom, Brian P and Paltiel, Yossi and Naaman, Ron and Waldeck,
               David H",
  journal   = "Chem. Rev.",
  publisher = "American Chemical Society (ACS)",
  volume    =  124,
  number    =  4,
  pages     = "1950--1991",
  month     =  "28~" # feb,
  year      =  2024,
  doi       = "10.1021/acs.chemrev.3c00661"
}

@ARTICLE{Yang2021-ld,
  title     = "Chiral spintronics",
  author    = "Yang, See-Hun and Naaman, Ron and Paltiel, Yossi and Parkin,
               Stuart S P",
  journal   = "Nat. Rev. Phys.",
  publisher = "Springer Science and Business Media LLC",
  volume    =  3,
  number    =  5,
  pages     = "328--343",
  month     =  "8~" # apr,
  year      =  2021,
  doi       = "10.1038/s42254-021-00302-9"
}

@ARTICLE{Menichetti2023-ob,
  title         = "Giant chirality-induced spin polarization in twisted
                   transition metal dichalcogenides",
  author        = "Menichetti, Guido and Cavicchi, Lorenzo and Lucchesi,
                   Leonardo and Taddei, Fabio and Iannaccone, Giuseppe and
                   Jarillo-Herrero, Pablo and Felser, Claudia and Koppens, Frank
                   H L and Polini, Marco",
  journal       = "arXiv [cond-mat.mes-hall]",
  month         =  "14~" # dec,
  year          =  2023,
  archivePrefix = "arXiv",
  primaryClass  = "cond-mat.mes-hall",
  doi           = "10.48550/arXiv.2312.09169"
}

@ARTICLE{Inbar2023-pp,
  title     = "The quantum twisting microscope",
  author    = "Inbar, A and Birkbeck, J and Xiao, J and Taniguchi, T and
               Watanabe, K and Yan, B and Oreg, Y and Stern, Ady and Berg, E and
               Ilani, S",
  journal   = "Nature",
  publisher = "Springer Science and Business Media LLC",
  volume    =  614,
  number    =  7949,
  pages     = "682--687",
  month     =  "22~" # feb,
  year      =  2023,
  doi       = "10.1038/s41586-022-05685-y"
}

@ARTICLE{Jayasena2015-bn,
  title     = "An investigation of {PDMS} stamp assisted mechanical exfoliation
               of large area graphene",
  author    = "Jayasena, Buddhika and Melkote, Shreyes N",
  journal   = "Procedia Manuf.",
  publisher = "Elsevier BV",
  volume    =  1,
  pages     = "840--853",
  year      =  2015,
  doi       = "10.1016/j.promfg.2015.09.073"
}

@ARTICLE{Bistritzer2010-md,
  title     = "Transport between twisted graphene layers",
  author    = "Bistritzer, R and MacDonald, A H",
  journal   = "Phys. Rev. B Condens. Matter Mater. Phys.",
  publisher = "American Physical Society (APS)",
  volume    =  81,
  number    =  24,
  pages     =  245412,
  month     =  "8~" # jun,
  year      =  2010,
  doi       = "10.1103/physrevb.81.245412"
}

@ARTICLE{Li2022-sa,
  title     = "Stacking and twisting of freestanding complex oxide thin films",
  author    = "Li, Ying and Xiang, Cheng and Chiabrera, Francesco M and Yun,
               Shinhee and Zhang, Haiwu and Kelly, Daniel J and Dahm, Rasmus T
               and Kirchert, Charline K R and Cozannet, Thomas E Le and Trier,
               Felix and Christensen, Dennis V and Booth, Timothy J and
               Simonsen, Søren B and Kadkhodazadeh, Shima and Jespersen, Thomas
               S and Pryds, Nini",
  journal   = "Adv. Mater.",
  publisher = "Wiley",
  volume    =  34,
  number    =  38,
  pages     = "e2203187",
  month     =  sep,
  year      =  2022,
  doi       = "10.1002/adma.202203187"
}

@ARTICLE{Pryds2024-wi,
  title     = "Twisted oxide membranes: A perspective",
  author    = "Pryds, N and Park, D-S and Jespersen, T S and Yun, S",
  journal   = "APL Mater.",
  publisher = "AIP Publishing",
  volume    =  12,
  number    =  1,
  pages     =  010901,
  month     =  "1~" # jan,
  year      =  2024,
  doi       = "10.1063/5.0176307"
}

@ARTICLE{Koren2016-ba,
  title     = "Coherent commensurate electronic states at the interface between
               misoriented graphene layers",
  author    = "Koren, Elad and Leven, Itai and Lörtscher, Emanuel and Knoll,
               Armin and Hod, Oded and Duerig, Urs",
  journal   = "Nat. Nanotechnol.",
  publisher = "Springer Science and Business Media LLC",
  volume    =  11,
  number    =  9,
  pages     = "752--757",
  month     =  sep,
  year      =  2016,
  doi       = "10.1038/nnano.2016.85"
}

@ARTICLE{Luican2011-hd,
  title     = "Single-layer behavior and its breakdown in twisted graphene
               layers",
  author    = "Luican, A and Li, Guohong and Reina, A and Kong, J and Nair, R R
               and Novoselov, K S and Geim, A K and Andrei, E Y",
  journal   = "Phys. Rev. Lett.",
  publisher = "American Physical Society (APS)",
  volume    =  106,
  number    =  12,
  pages     =  126802,
  month     =  "25~" # mar,
  year      =  2011,
  doi       = "10.1103/PhysRevLett.106.126802"
}

@ARTICLE{Kim2013-al,
  title     = "Breakdown of the interlayer coherence in twisted bilayer graphene",
  author    = "Kim, Youngwook and Yun, Hoyeol and Nam, Seung-Geol and Son,
               Minhyeok and Lee, Dong Su and Kim, Dong Chul and Seo, S and Choi,
               Hee Cheul and Lee, Hu-Jong and Lee, Sang Wook and Kim, Jun Sung",
  journal   = "Phys. Rev. Lett.",
  publisher = "American Physical Society (APS)",
  volume    =  110,
  number    =  9,
  pages     =  096602,
  month     =  "1~" # mar,
  year      =  2013,
  doi       = "10.1103/PhysRevLett.110.096602"
}

@ARTICLE{Yang2020-es,
  title     = "In situ manipulation of van der Waals heterostructures for
               twistronics",
  author    = "Yang, Yaping and Li, Jidong and Yin, Jun and Xu, Shuigang and
               Mullan, Ciaran and Taniguchi, Takashi and Watanabe, Kenji and
               Geim, Andre K and Novoselov, Konstantin S and Mishchenko, Artem",
  journal   = "Sci. Adv.",
  publisher = "American Association for the Advancement of Science (AAAS)",
  volume    =  6,
  number    =  49,
  pages     = "eabd3655",
  month     =  dec,
  year      =  2020,
  doi       = "10.1126/sciadv.abd3655"
}

@ARTICLE{Naaman2012-ow,
  title     = "Chiral-induced spin selectivity effect",
  author    = "Naaman, R and Waldeck, David H",
  journal   = "J. Phys. Chem. Lett.",
  publisher = "American Chemical Society (ACS)",
  volume    =  3,
  number    =  16,
  pages     = "2178--2187",
  month     =  "16~" # aug,
  year      =  2012,
  doi       = "10.1021/jz300793y"
}

@ARTICLE{Chari2016-gl,
  title     = "Resistivity of rotated graphite-graphene contacts",
  author    = "Chari, Tarun and Ribeiro-Palau, Rebeca and Dean, Cory R and
               Shepard, Kenneth",
  journal   = "Nano Lett.",
  publisher = "American Chemical Society (ACS)",
  volume    =  16,
  number    =  7,
  pages     = "4477--4482",
  month     =  "13~" # jul,
  year      =  2016,
  doi       = "10.1021/acs.nanolett.6b01657"
}

@ARTICLE{Francesco-M-Chiabrera-Shinhee-Yun-Ying-Li-Rasmus-T-Dahm-Haiwu-Zhang-Charline-K-R-Kirchert-Dennis-V-Christensen-Felix-Trier-Thomas-S-Jespersen-and-Nini-Pryds2022-jq,
  title    = "Freestanding Perovskite Oxide Films: Synthesis, Challenges, and
              Properties",
  author   = "{Francesco M. Chiabrera, Shinhee Yun, Ying Li, Rasmus T. Dahm,
              Haiwu Zhang, Charline K. R. Kirchert, Dennis V. Christensen, Felix
              Trier, Thomas S. Jespersen, and Nini Pryds}",
  journal  = "Ann. Phys.",
  volume   =  534,
  pages    =  202200084,
  year     =  2022,
  doi      = "10.1002/andp.202200084"
}

@ARTICLE{Lee2015-pb,
  title     = "Emergence of room-temperature ferroelectricity at reduced
               dimensions",
  author    = "Lee, D and Lu, H and Gu, Y and Choi, S-Y and Li, S-D and Ryu, S
               and Paudel, T R and Song, K and Mikheev, E and Lee, S and
               Stemmer, S and Tenne, D A and Oh, S H and Tsymbal, E Y and Wu, X
               and Chen, L-Q and Gruverman, A and Eom, C B",
  journal   = "Science",
  publisher = "American Association for the Advancement of Science (AAAS)",
  volume    =  349,
  number    =  6254,
  pages     = "1314--1317",
  month     =  "18~" # sep,
  year      =  2015,
  doi       = "10.1126/science.aaa6442"
}

@ARTICLE{Imada1998-mp,
  title     = "Metal-insulator transitions",
  author    = "Imada, Masatoshi and Fujimori, Atsushi and Tokura, Yoshinori",
  journal   = "Rev. Mod. Phys.",
  publisher = "American Physical Society (APS)",
  volume    =  70,
  number    =  4,
  pages     = "1039--1263",
  month     =  "1~" # oct,
  year      =  1998,
  doi       = "10.1103/revmodphys.70.1039"
}

\end{document}